

\def\eps{\varepsilon}
\def\ii{\'{\char'20}}
\baselineskip=15pt plus 1pt minus 2pt 
\font\fss=cmss10 at 10pt \font\flbf=cmbx10 at 12pt
\parskip=10pt plus 1pt minus 2pt
\def\toda{\number\day \space\ifcase\month\or Jan\or Feb\or Mar\or Apr\or
May\or Jun\or Jly\or Aug\or Sep\or Oct\or Nov\or Dec\fi \space\number\year}
\global\newcount\eqn  \global\eqn=0
\global\newcount\sec \global\sec=0
\global\newcount\ftno \global\ftno=0
\def\roman#1{\ifcase#1 O\or I\or II\or III\or IV\or V\or VI\or VII\else
              ?? \fi}
\def\neq{\global\advance\eqn by1 \eqno(\the\sec . \the\eqn)}
\def\section#1{\global\advance\sec by1 \bigskip \bigskip \noindent
{\flbf \roman{\sec}.-  #1} \bigskip \global\eqn=0}
\def\label#1{\xdef#1{\hbox{(\the\sec.\the\eqn)~}}}
\def\ftnote#1{\global\advance\ftno by1 \footnote{$^{\the\ftno}$}{#1}}
\def\frac#1#2{{#1 \over #2}}
\def\sqr#1#2{{\vcenter{\hrule height.#2pt \hbox{\vrule width.#2pt
height#1pt \kern#1pt \vrule width.#2pt} \hrule height.#2pt}}}
\def\sq{{\mathchoice\sqr55\sqr55\sqr{2.1}3\sqr{1.5}3}\hskip 1.5pt}
\def\lrhup#1{\buildrel {{\leftharpoonup \hskip -8pt \rightharpoonup}}
\over #1}
%
%
\global\newcount\refno \global\refno=0
\newwrite\rfile
\def\ref#1#2{\global\advance\refno by1 \xdef#1{[\the\refno]}
\ifnum\refno=1\immediate\openout\rfile=refs.aux\fi
\immediate\write\rfile{\noexpand \item{[\the\refno]}{#2}}}


\ref\bogoliubov{N.N. Bogoliubov, O.S. Parasiuk, Acta. Math. {\bf
97} 227 (1957);  O.S. Parasiuk, Ukr. Math. Z. {\bf 12}, 287 (1960);
N.N. Bogoliubov, O.V. Shirkov, ``Introduction to the Theory of Quantized
Fields", 4th edition, Wiley, New York (1980).}

\ref\hepp{K. Hepp, Comm. Math. Phys. {\bf 2}, 301 (1966);  K. Hepp, La
``Th\'eorie de la Renormalisation", Lecture Notes in Physics {\bf 2},
Springer Verlag, Berlin (1969).}

\ref\zimm{W. Zimmermann, Comm. Math. Phys. {\bf 15}, 208 (1969);  {\bf 11},
1 (1968).}

\ref\dyson{F.J. Dyson, Phys. Rev. {\bf 75}, 486 (1949); {\bf 75}, 1736
(1949).}

\ref\salam{A. Salam, Phys. Rev. {\bf 82}, 217 (1951); {\bf 84} 426
(1951).}

\ref\stuckgreen{E.C.G. Stuckelberg, T.A. Green, Helv. Phys. Acta. {\bf 24},
153 (1951).}

\ref\weinberg{S. Weinberg, Phys. Rev. {\bf 118}, 838 (1960).}

\ref\epstein{H. Epstein and V. Glaser, Ann.Inst.Henri Poincare {\bf XIX},
211 (1973);  H. Epstein and V. Glaser in ``Statistical Mechanics and
Quantum
Field Theory", {\sl Proceedings of the Les Houches Summer School 1970},
C. de Witt and R. Stora eds, Gordon and Breach, New York (1971).}

\ref\callan{C.G. Callan, Phys. Rev. {\bf D2}, 1541 (1970);  A.S. Blaer, K.
Young, Nucl. Phys. {\bf B83}, 493 (1974);  C.G. Callan in ``Methods in
Field
Theory", {\sl Proceedings of the Les Houches Summer School 1975}, R.
Balian and J. Zinn-Justin eds., North-Holland, Amsterdam (1976).}

\ref\polchinski{J. Polchinski, Nucl. Phys. {\bf B231,\/} 269 (1984).}

\ref\fjl{D.Z. Freedman, K. Johnson, J.I. Latorre,
Nucl. Phys. {\bf B371,\/} 353 (1992).}

\ref\peter{P.E. Haagensen, Mod. Phys. Lett. {\bf A7,\/} 893 (1992).}

\ref\threegluon{D.Z. Freedman, G. Grignani, K. Johnson, N. Rius,
Ann. Phys. (N.Y.) {\bf 218, \/} 75 (1992)}

\ref\massive{P.E. Haagensen, J.I. Latorre,
Phys. Lett. {\bf B283, \/} 293 (1992).}

\ref\ramon{R. Mu\~noz-Tapia, Phys. Lett. {\bf B295}, 95 (1992).}

\ref\qed{P.E. Haagensen, J.I. Latorre, Ann. Phys. (N.Y.) {\bf 221}, 77
(1993).}

\ref\cris{C. Manuel, Niels Bohr Institute preprint NBI-HE-92-74, to
appear in Int. J. Mod. Phys. {\bf A}.}

\ref\dunnerius{G. Dunne, N. Rius, Phys. Lett. {\bf B293}, 367 (1992).}

\ref\stonybrook{D.Z. Freedman, ``Differential regularization and
renormalization: recent progress", in {\sl Proceedings of the
Stony Brook Conference on Strings and Symmetries,\/} Spring, 1991.}

\ref\FJMV{D.Z. Freedman, K. Johnson, R. Mu\~noz-Tapia, X.
Vilas\ii s-Cardona, ``A Cutt Off Procedure and Countertemrs for
Differential Renormalization", to appear in Nucl. Phys. {\bf B}}

\ref\lowenstein{J.H. Lowenstein, ``Seminars on Renormalization Theory",
vol. II., University of Maryland Technical Report \#73-068.}

\ref\itzzuber{C. Itzykson, J.B. Zuber, ``Quantum Field Theory",
McGraw-Hill, New York (1980).}

\ref\collins{J.C. Collins, ``Renormalization", Cambridge University
Press, Cambridge (1984).}

\ref\bonneau{G. Bonneau, Int. J. Mod. Phys. {\bf A5}, 3831 (1990).}

\ref\smith{W.E. Caswell and A.D. Kennedy, Phys. Rev. {\bf D25}, 329
(1982).}

\ref\gouyon{R. Gouyon, ``Int\'egration et distributions", Librairie
Vuibert, Paris (1979).}

\ref\lowens{J.H. Lowenstein, M. Weinstein, W. Zimmermann, Phys. Rev.
{\bf D10}, 1854 (1974); Phys. Rev. {\bf D10}, 2500 (1974).}

\font\fvlbf=cmbx10 at 17.28pt
\pageno=0
\footline={}
\null
\vfill
\centerline{\fvlbf Systematic Differential Renormalization}
\centerline{\fvlbf to All Orders}
\bigskip \bigskip
\centerline{\flbf
Jos\'e I. Latorre\footnote{*}{{\rm bitnet : }{\tt latorre@ebubecm1}},
Cristina Manuel\footnote{**}{{\rm bitnet : }{\tt palas@ebubecm1}},
Xavier Vilas\ii s-Cardona\footnote{***}{{\rm bitnet : }{\tt
druida@ebubecm1}}} \bigskip
\centerline{\it Departament d'Estructura i Constituents de la Mat\`eria}
\centerline{\it Facultat de F\ii sica, Universitat de Barcelona}
\centerline{\it Diagonal 647, 08028-Barcelona\ \ SPAIN}

\vfill
\centerline{\flbf Abstract}

{\baselineskip=13pt
We present a systematic implementation of differential renormalization
to all orders in perturbation theory.   The method is applied to
individual Feynman graphs written in coordinate space. After
isolating every  singularity which appears in a bare diagram, we
define a subtraction procedure which consists in  replacing
the core of the
singularity by its renormalized form given by a differential formula.
The organization of subtractions in subgraphs relies on
Bogoliubov's formula, fulfilling the requirements of locality,
unitarity and Lorentz invariance. Our method bypasses the use
of an intermediate regularization and  automatically delivers
renormalized amplitudes which obey renormalization group equations.}

\vfill
\noindent{\bf UB-ECM-PF 93/4 \hfill  \toda}
\eject

\footline{\hss \folio \hss}

\section{Introduction}

It is well known that the amplitudes of the perturbative expansion
of an interacting quantum field theory have, in general, an ill-defined
ultraviolet
behavior. From a mathematical point of view, the core of this problem
lies on the nature of these amplitudes, which are distribution-valued
objects, since the product of distributions is known to be, in general,
ill-defined. As stated in ref.\bogoliubov ,
renormalization consists, thus, in finding a prescription to define the
product of distributions so that amplitudes verify some desired
requirements, namely, Lorentz invariance, locality and unitarity.
This problem was shown to have a solution long ago. Among the different
ideas to prove the existence of a consistent renormalization
program, let us
single out the approach started by Bogoliubov and Parasiuk \bogoliubov~
and definitively settled by Hepp \hepp~and Zimmermann \zimm
\ftnote{Many people have contributed to set up the basis of
renormalization theory. Among many others let us mention the works of
Dyson \dyson, Salam \salam, Stuckelberg and Green \stuckgreen, Weinberg
\weinberg, Epstein and Glaser \epstein, Callan, Blaer and
Young \callan~ and Polchinski \polchinski.}.
The
BPHZ (Bogoliubov, Parasiuk, Hepp and Zimmermann) method is based on the
concept of counterterms and defines a recursive subtraction scheme which
can be applied on individual Feynman graphs.
{}From these works, we learn that the renormalization program proceeds as
follows. Locate first the divergences occuring in the bare amplitude of
the studied graph,
apply then a subtraction procedure, that is, a method to eliminate
such divergences, and finally organize the subtraction according to the
topology of the graph.
In this paper,
 we define a differential
renormalization subtraction procedure to be
implemented  in  the program described above.

Let us recall the simple principles of differential renormalization
(DR), as presented in \fjl.
The procedure is defined in coordinate space and yields right away
renormalized amplitudes. Divergent expressions
are written as derivatives of less singular functions.
We have then to solve a differential equation, extracting as many
derivatives as necessary to obtain
a power-counting-finite expression. The differential equation is
promoted to be
a definition of the amplitude in the sense of distributions. Derivatives
are, in this context, naturally understood to act on test
functions. When solving those differential equations, integration
constants appear, which play the role of renormalization group scales.
Differential renormalization amplitudes
fulfill renormalization group equations, which are used to extract the
coefficients
of the perturbative expansion of renormalization group $\beta$ and
$\gamma$ functions. The main advantages of the
method are the relative ease of the calculations and the fact that
the space-time dimension remains unchanged.
It has been successfully
applied to massless $\phi^4$ to three loops in four dimensions \fjl,
supersymmetric theories \peter, the study of the three-gluon vertex
\threegluon, massive theories \massive, low dimensional theories
\ramon, QED to two loops \qed~and theories with $\gamma_5$ \cris.
It has also been related to the standard dimensional regularization
procedure in reference \dunnerius.

Our purpose is
to fill the main gap of differential renormalization, namely, the lack
of a systematic set of rules. In spite of its successes and
achievements, one can doubt whether differential
 renormalization principles are
sufficient to render finite {\it any} Feynman graph. Moreover,
without a systematic procedure, it is
impossible to prove that differential renormalization
amplitudes are consistent and
unitary to all orders. By consistent, we understand following from a
counterterm
structure or fulfilling renormalization group equations, which is
equivalent. At the present
level of differential renormalization, we cannot ensure, for instance,
that
overlapping divergences are treated correctly. Although differential
renormalization amplitudes
obey renormalization group equations at a given order
(which are used to compute $\beta$ and
$\gamma$ renormalization group-functions), this is not sufficient to
guarantee a counterterm structure since we can
think of amplitudes fulfilling renormalization group
equations up to a certain order in perturbation
theory and yet not being consistent \stonybrook.
An attempt to clarify the existence of a counterterm
structure behind
differential renormalization amplitudes was
made in reference \FJMV. By introducing a cut off $\eps$ in massless
four-dimensional $\phi^4$ theory to three loops, one finds that
divergences organize correctly in counterterms, yielding
the differential renormalization
renormalized amplitudes. However, a check, no matter
how thorough, is never a proof. Therefore, for the sake of completeness,
a systematization of differential renormalization is needed. Then, the
consistency and unitarity of the renormalized amplitudes can be checked
to all orders.

We can summarize the general idea of our approach as follows.
The differential renormalization procedure is defined in coordinate
space. We distinguish between divergences arising from two points
collapsing and those coming from three or more points simultaneously
closing up. We define for the first our particular subtraction
procedure which consists of applying the basic idea of ``pulling out"
derivatives. We actually organize the subtraction as a true replacement
of the singularity with the renormalized form once the
derivatives are pulled in front.
 For the second, we work recursively by
first observing that no singularity appears when all points but
one are brought together. This indicates that the global
singularity reduces to the problem of bringing the last
point on top of the rest, which is just a two-point problem
again. This simplification is essentially due to the fact that
the subtraction to be performed is local.
 No regulator is needed to define
these subtractions, sharing one of the characteristic features of the
BPHZ scheme. The subtraction
of subdivergences is organized following Bogoliubov's recursion formula.
We therefore come up with a systematic version of
differential renormalization which guarantees the
desired properties of consistency and unitarity of the method.
The expert reader should be aware of the fact that  we do not
attempt to present an exhaustively rigorous proof.
It is a virtue of DR to
remain an extremely simple method to apply, regardless of all the
technicalities we are borrowing to prove its workability to all orders.

The organization of our paper goes as follows. In section II, we first
recall the basic ideas of the renormalization proof adapting them to
our coordinate space approach.
We then sketch two examples of how subtractions will be defined
and proceed to fully present the general procedure.
 Section III is
devoted to non-trivial illustrations of the method, going from a
three-loop non-planar diagram to a six-loop one.
We end up with a series of comments
on the differences between our method and the standard BPHZ, and its
applicability  to theories with supplementary symmetries. Two appendices
are devoted to technical questions.

\section{Systematic Differential Renormalization}

{\bf 1.- Organization of the Renormalization Procedure}

A correct renormalization procedure must provide a method to define
sensible Green functions while preserving locality, unitarity and
Lorentz invariance. In momentum space, Green functions are
ill-defined when loop integrals blow up at high momenta.
Therefore, renormalization amounts to eliminating those
infinities appearing in loop integrals. In coordinate space,
divergences arise when vertices come close. We have then
to smear out such divergent behavior
so that amplitudes end up being (tempered) distributions.
As we have already said, this program can be achieved by a three-step
process. Let us briefly comment on each step, with a preview of how
differential renormalization will adapt to them.

Given a Feynman graph, we have first to know whether it needs
renormalization.
Relying on Weinberg's theorem \weinberg, power counting techniques
provide the tools for this job.
Weinberg's theorem states that a graph which is power
counting finite and whose subgraphs are also power counting finite is
finite. Therefore, to locate potential divergences in coordinate space,
we must investigate the superficial degree of divergence of every set of
vertices closing up.  Once a divergence is detected we have to obtain
its precise form. For instance, the method of BPHZ would instruct us to
expand the loop
integrands in Taylor series of the external momenta. Instead, we
propose a simple coordinate space method to find an equivalent
divergence which depends only on two points but has the same
relevant singular behavior.

The following step is to define a subtraction procedure to cure the
divergences from the bare amplitude. For instance, BPHZ
proceeds by eliminating the
divergent terms in the Taylor expansion used in the first step.
(If, alternatively, we were using the minimal subtraction
scheme in dimensional regularization we would analytically
continue the loop integrals in terms of the space-time dimension,
expand them in Laurent series around the pole and substract the
pole). In any case, such
subtractions must be local for primitively divergent graphs in order to
preserve locality. We recognize a local subtraction in momentum space by
being polynomial in the external momenta. In coordinate space a local
subtraction has  support only on the coinciding vertices. We
actively make use of this last observation and construct a recursive
subtraction of the divergences isolated in step one. Each subtraction is
done through a differential formula in such a way that the renormalized
form of the initial singularity is delivered.

Different subtraction procedures define different renormalization
procedures. However, they must all organize the subtraction of the
subdivergences of a given Feynman graph according to its topology
(see for instance references \lowenstein, \itzzuber,
\collins, \bonneau~and \smith), following Bogoliubov's recursion formula
\bogoliubov \hepp, in order to eliminate all such subdivergences and
still preserve locality. This is proved by induction in the number of
loops showing that, if we can consistently eliminate all subdivergences
up to a given order, the remaining overall divergence coming out at the
next order is local. A particular solution of Bogoliubov's formula
was given by Zimmermann \zimm, who defined the well known forests of
renormalization. This formulation can easily be shown
to come from a counterterm structure \bogoliubov \smith, provided the
subtraction procedure is local for primitively divergent graphs. Recall
that, when divergences organize in a counterterm structure, they can be
absorbed in the parameters of the theory (fields and couplings
constants). From counterterms we define bare fields and couplings.
Divergences are then cancelled when bare fields and couplings are
written in terms of physical quantities. In this picture, the
renormalization scale appears to be a new parameter needed to separate
divergent from finite parts, so bare amplitudes are independent of it.
This statement of independence leads to renormalization group equatio
ns for renormalized amplitudes.
RGE show that changes in the renormalization scale are also absorbed
into redefinitions of the couplings and fields. Since the form of the
lagrangian remains the same, Lorentz invariance is preserved. On the
other hand, if counterterms are hermititian, the lagrangian maintains,
at least formally, its hermiticity and, thus, the $S$-matrix remains
unitary.

Let us summarize this brief review. A subtraction procedure which is
local for primitively divergent graphs and which is implemented into
Bogoliubov's formula, ensures the renormalization of any Feynman diagram
and the existence of a counterterm structure, which, if hermititian,
ensures the unitarity of renormalized amplitudes.
Such procedure is therefore a correct renormalization procedure.

{\bf 2.- Basic Examples of the Differential Renormalization
Subtraction Procedure}

Let us start by recalling the differential renormalization (DR)
techniques in two illustrative
examples of four-dimensional Euclidean massless $\lambda \phi^4$. From
them, we devise a subtraction operation that yields the same DR
amplitudes and preserves Euclidean invariance and locality.

The 1PI one-loop four point amplitude of Euclidean massless
$\lambda \phi^4$, see {\fss Fig.1}, is \ftnote{Throughout this paper
we use the notation $\delta (x) \equiv \delta^{(4)} (x)$ and $x^2=
x_{\mu} x^{\mu}$.}
 $$
\Gamma^{bare} (x,y,z,w) = \frac{\lambda^2}{2} \delta (x-z)
\delta(y-w) \left ( \Delta (x-y) \right)^2 + (2-{\rm perm.}),
\neq \label\bbubble
$$
where
$$
\Delta (x) ={1\over 4\pi^2x^2},
\neq \label\propagator
$$
is the propagator. We can
set $y=0$, due to translational invariance. Even if the propagator
\propagator is a well-defined distribution, its square is not. This
problem manifests itself in the fact that the factor $1/x^4$ upon
Fourier transformation produces a logarithmic divergence.
 To treat this divergence following DR  one has
to solve the differential equation
$$
\frac {1}{x^4} = \sq A (x).
\neq \label\diffeq
$$
The solution of Eq. \diffeq is
$$
\frac {1}{x^4} = -\frac{1}{4} \sq \frac {\ln x^2 M^2}{x^2}.
\neq \label\diffbubble
$$
The degree of divergence of the solution of Eq.\diffeq
has been reduced by two, as a naive power counting shows.
Since the bare amplitude is logarithmically divergent,
it would have been sufficient to consider a first order
differential equation.
The degree of divergence would then have been reduced by one, yielding a
correct renormalized factor.
 However, the use of the laplacian instead of a linear
derivative allows to easily  impose manifest Euclidean invariance on
the solution of Eq.\diffeq
just by requiring that $A$ is a function of $x^2$.
This increasing of the natural order of the differential equation yields
an additive integration constant, dropped to ensure
sensible power damping of amplitudes at infinity. In any case,
the Fourier transform of this constant is a delta function in
momentum space which vanishes as the laplacian produces powers of
momenta when it acts by parts.
 The other integration constant, $M$, is a
mass scale that plays a central role in the method: it is the
renormalization scale of the amplitude. We will come back to this issue
later on.

It is important to note that  equality Eq.\diffbubble is exact for
all values of $x$ except
for $x=0$, where it is undefined. The fact is that the
right-hand-side
of \diffbubble is a well-defined object in the sense of distributions.
 We can define thus
a DR subtraction operator $T$ that allows to eliminate the
strong singular behavior at short distances of bare amplitudes, without
altering them in any other region, giving
$$
\Gamma ^{ren} = (1 - T) \Gamma ^{bare},
\neq \label\Tsubtr
$$
where $\Gamma ^{ren}$ is now a well-defined distribution.
Following this formulation,
the operation that selects the singular part of the
$1/x^4$ factor is
$$
T_{x,0} \frac {1}{x^4} \equiv
\frac {1}{x^4} -\left ( -\frac{1}{4} \sq \frac {\ln x^2 M^2}{x^2}
\right),
\neq \label\Tbubble
$$
so that using Eq.\Tsubtr the correct DR renormalized
amplitude can be obtained. Introducing a regulator in Eq.\Tbubble, as
done in ref. \FJMV, we can give an explicit meaning to the formal
operation carried out here. Recall that the regulated progator is
$$
\Delta_\epsilon (x) = \frac1{4\pi^2} {1 \over x^2 + \epsilon^2},
\neq
$$
and we have
$$
{1\over x^4} \rightarrow {1\over (x^2 + \epsilon^2)^2} = -\frac14 \sq
{\ln (x^2+\epsilon^2) /\epsilon^2 \over x^2} = -\frac14 \sq {\ln
(x^2+\epsilon^2)M^2 \over x^2} -\pi^2\ln \epsilon^2 M^2 \ \ \delta(x).
\neq
$$
Therefore, the subtraction is
$$
T_{x,0}^{\epsilon} \frac {1}{(x^2+ \epsilon^2)^2} =
-\pi^2\ln \epsilon^2 M^2 \ \ \delta(x).
\neq \label\Trbubble
$$
The use of a regulator is, however, not necessary. The $T$ operator
isolates the singular part of the factor $1/x^4$, which only
has support in $x=0$.
It can be subtracted using Eq.\Tsubtr and render the amplitude
renormalized. The requirement of locality is in this way
preserved. We also note that the operator $T$ is obviously shaped as a
replacement. It substitutes a singular expression by its renormalized
form.

The DR renormalized amplitude corresponding to Eq.\bbubble can be
finally written as
$$
\Gamma^{ren} (x,y,z,w) = -\frac{\lambda^2}{128 \pi^4} \left (
\delta (x-z) \delta (x-w) \sq \frac {\ln (x-y)^2 M^2}{(x-y)^2} +
(2-{\rm perm.}) \right).
\neq \label\rbubble
$$
Notice that the dependence of Eq.\rbubble on the mass scale $M$
is such that
$$
M \frac{\partial}{\partial M} \Gamma^{ren} (x,y,z,w) =
\frac {3 \lambda^2}{16 \pi^2} \delta(x-z) \delta(x-w) \delta(x-y),
\neq \label\massd
$$
so a change in $M$ can be reabsorbed in a change of the coupling
constant $\lambda$. This is an important property of DR. DR
renormalized amplitudes automatically satisfy renormalization group
equations with $M$ playing the role of  renormalization scale. From them
we extract the $\beta$ function, obtaining
$$
\beta = \frac {3\lambda}{16 \pi^2},
\neq \label\beta
$$
which is the correct result.

We center now our attention on the study of a two-loop bare
amplitude, ({\fss Fig.2}) given by
$$
\Gamma^{bare} (x,y,z,w) = -\frac{\lambda^3}{2} \left (
\delta (z-w) \Delta(x-z) \Delta(y-z) \left ( \Delta (x-y) \right)^2
+ {\rm (5-perm.)} \right).
\neq \label\bicecone
$$
The singular factor of Eq.\bicecone
is
 $$
f(x,y,0) = \frac{1}{x^2y^2}\frac{1}{(x-y)^4},
\neq \label\bicefactor
$$
(we have set $z=0$). One first has to treat the subfactor
$1/(x-y)^4$ in the same way that it was done in Eq.\diffbubble,
so
 $$
\left (1 - T_{x,y} \right ) f(x,y,0)
 =  -\frac{1}{4}
\frac {1}{x^2 y^2} \sq \frac {\ln(x-y)^2 M^2}{(x-y)^2}.
\neq \label\icecone
$$
After curing the subdivergence $x \sim y$, $f$ is still
logarithmically divergent when the three points, $x,y,0$,  approach
each other simultaneously. However, notice that now every factor of
\icecone, $1/x^2$, $1/y^2$ and $\sq \ln (x-y)^2M^2 / (x-y)^2$, is a
well-defined object in the sense of distributions.
Only their product is not.
In the spirit
of DR, one should define the renormalized factor corresponding to Eq.
\icecone as derivatives of a less
singular function. A priori it is not clear how one should proceed here,
since there are two independent variables, and it is not obvious how
derivatives should be extracted. The strategy followed in ref.\fjl~
was to
move the laplacian to the left, by some exact manipulations, in order to
use
$$
\sq \frac{1}{x^2} = - 4\pi^2 \delta (x),
\neq \label\deltta
$$
The problem was then reduced basically to that of
one variable, and the principle of extracting derivatives could be
easily implemented. However, the use of \deltta is very much dependent
on the propagator factors appearing in the amplitude. Eq. \deltta cannot
be applied to some higher loop graphs
(see the example of III.3). We will generalize
the subtraction procedure that we used in the previous case, where
problems arose when two points came close, to this case, where we deal
with a three-point problem.

Notice, first, that Eq.\icecone and
$$
\delta(y) \int d^4y
\frac {1}{x^2 y^2} \sq \frac {\ln(x-y)^2 M^2}{(x-y)^2}
= -4\pi^2 \delta (y) \frac {\ln x^2M^2}{x^4},
\neq \label\ice
$$
have the same divergent behavior when $x\sim y\sim 0$, therefore they
will need the same subtraction. To prove it, notice that \icecone
is a well-defined distribution in $y$ and study
$$
\int d^4y \,\varphi (y)
\frac {1}{x^2 y^2} \sq \frac {\ln(x-y)^2 M^2}{(x-y)^2},
$$
where $\varphi (y)$ is a test function. Let us write,
$$
\int d^4y \,\left( \varphi (y) - \varphi (0) \right)
\frac {1}{x^2 y^2} \sq \frac {\ln(x-y)^2 M^2}{(x-y)^2} +
\varphi (0) \int d^4y \,
\frac {1}{x^2 y^2} \sq \frac {\ln(x-y)^2 M^2}{(x-y)^2}.
\neq
$$
The second term corresponds to the action of \ice onto the test
function. The factor $\left( \varphi (y) - \varphi (0) \right)$ is a
test function which vanishes in the origin. We can write it as
$\vert y \vert \psi (y)$, being $\psi (y)$ a test function. The first
term will therefore yield a well defined distribution in $x$ since the
factor
$$
\frac {\vert y \vert}{x^2 y^2} \sq \frac {\ln(x-y)^2 M^2}{(x-y)^2}
$$
is now power counting finite. Thus, we have concentrated the divergent
behavior in $x$ around the point 0 in the second term. Therefore
the divergent behavior of Eq.\icecone when $x \sim y \sim 0$ is in
effect given by Eq.\ice. In appendix A an alternative proof
of this statement can be found. Let us remark that all the integrals
considered are well behaved in the infrared region.

We want to find a subtraction that renormalizes \icecone and is local,
in other words, with support only on $x=y=0$. Since \icecone and \ice
have the same behavior in the conflictive region $x\sim y\sim 0$, they
will both need the same subtraction. In this way, we
have reduced a three-point problem to that of a two-point function.
We can now use the principle of extracting derivatives described in the
previous example, so we have to solve the following differential
equation
$$
\frac{\ln x^2M^2}{x^4} = \sq B(x).
\neq \label\diffloop
$$
The solution of Eq.\diffloop is
$$
\frac{\ln x^2M^2}{x^4} =
-\frac{1}{8} \sq \frac {\ln^2 x^2M^2 + 2 \ln x^2M'^{2}} {x^2},
\neq \label\twoloop
$$
where $M'$ is an integration constant that has dimensions of mass. The
other integration constant present in the general solution of
Eq.\diffloop has been dropped for the same reasons as
in Eq.\diffbubble.

The $T$-operator that subtracts the singular part of Eq.\icecone
is in this case
$$
T_{x,y,0} \left(1-T_{x,y} \right) f(x,y,0) \equiv
\pi^2 \delta(y) \left ( \frac {\ln
x^2M^2}{x^4} + \frac {1}{8} \sq \frac {\ln^2 x^2M^2 + 2 \ln
x^2M'^2}{x^2} \right),
\neq \label\Ticecone
$$
so  that
$$
f^{ren} (x,y,0) = (1 - T_{x,y,0})\left(1-T_{x,y} \right) f(x,y,0) $$ $$
 =  -\frac{1}{4} \frac{1}{x^2y^2} \sq  \frac{\ln (x-y)^2M^2}{(x-y)^2}
- \pi^2 \delta (y) \left (  \frac {\ln x^2M^2}{x^4} +
\frac{1}{8} \sq \frac{\ln^2 x^2M^2 + 2 \ln x^2M'^2}{x^2}
\right).
\neq \label\renicecone
$$
The sum of these three terms is a well-defined distribution, even though
the first two are not. One can now perform
exact manipulations in the first term of Eq.\renicecone,
taking into account
the identity
$$
A \sq B = \partial_{\nu} \left (A
\lrhup{\partial_{\nu}} B \right) + B \sq A.
\neq \label\diffid
$$
Then, using formula Eq.\deltta, we have
$$
f^{ren} (x,y,0) =   $$ $$
= -\frac{1}{4} \frac{\partial}{\partial y_{\nu}} \left (
\frac{1}{x^2y^2}
{\frac{\lrhup{\partial}}{\partial y_{\nu}}}
 \frac{\ln (x-y)^2M^2}{(x-y)^2} \right) + \pi^2 \delta (y)
 \frac {\ln x^2M^2}{x^4} $$ $$
- \pi^2 \delta (y) \left (  \frac {\ln x^2M^2}{x^4} +
\frac{1}{8} \sq \frac{\ln^2 x^2M^2 + 2 \ln x^2M'^2}{x^2}
\right) $$ $$
= -\frac{1}{4} \frac{\partial}{\partial y_{\nu}} \left (
\frac{1}{x^2y^2}
{\frac{\lrhup{\partial}}{\partial y_{\nu}}}
 \frac{\ln (x-y)^2M^2}{(x-y)^2} \right)
- \frac{\pi^2}{8} \delta(y) \sq \frac{\ln^2 x^2M^2 + 2 \ln
x^2M'^2}{x^2}.
\neq
$$

Hence, the complete renormalized amplitude is
$$
 \Gamma^{ren} (x,y,z,w) =  $$ $$
=\frac {\lambda^3}{8(2\pi)^8} \left( \delta(x-w)
\frac{\partial}{\partial y_{\nu}} \left (
\frac{1}{(x-z)^2(y-z)^2}
{\frac{\lrhup{\partial}}{\partial y_{\nu}}}
 \frac{\ln (x-y)^2M^2}{(x-y)^2} \right) \right.  $$ $$
+  \frac{\pi^2}{2} \delta(x-w) \delta(y-z) \sq \frac{\ln^2
(x-z)^2M^2 + 2 \ln (x-z)^2M'^2}{(x-z)^2} + {\rm (5-perm)}
 \Bigg ).
\neq \label\renamplice
$$
This is the result obtained in ref.\fjl. It should be noted that
this amplitude in momentum space has a far more complicated expression
than the one given in Eq.\renamplice, since in momentum space it
contains a dilogarithm function.
 Notice  also that
we find two different mass scales in Eq.\renamplice, $M$ and
$M'$.
The first one, $M$, comes from the renormalization of the subdivergence,
and is the same that appears in the one-loop renormalized
amplitude Eq.\rbubble. Here it has been promoted from $\ln
x^2M^2$
to $\ln^2 x^2M^2$: this is nothing but the action of the renormalization
group. The new mass scale $M'$ has been introduced to renormalize the
overall divergence. One could set to any arbitrary value the
relation between $M$ and
$M'$, the final selection of this relation being fixed by the choice of
a particular renormalization scheme.

Up to now we have carefully
analyzed two amplitudes in the way DR processes them. We
have defined a subtraction operation which replaces by its renormalized
form the singular part of
those bare amplitudes for a two-point
and a three point problem, and we have shown  that this last case can be
reduced to the former. This is a very important feature that will be
used to
study  n-point primitively divergent graphs in the following subsection.
The keystone of this
reduction is the locality of the divergence occurring in such n-point
functions.

We would like to point out here an important issue. Formulas such as
Eq.\diffid were used in ref.\fjl~in order
to easily locate and identify divergent factors that needed
renormalization.
 These kind of manipulations are exact and thus were correctly
used in ref.\fjl. They are totally
harmless since they do not alter the singular behavior of bare
amplitudes. Such manipulations of derivatives should not be
confused with  the real core of DR, which lies
in the use of differential formulas, as
Eq.\diffbubble and Eq.\twoloop. DR promotes
 these differential formulas  to be the
 definitions of renormalized amplitudes in the sense of distributions.
The integration by parts rule is naturally implemented in this framework
and so reduces the degree of divergences of bare amplitudes.

{\bf 3.-General procedure}

We present in this subsection the general rule that DR prescribes to
process an arbitrary Feynman graph.
 For the sake of simplicity, we are
going to consider massless scalar theories without derivative couplings,
 working always in Euclidean space.

We start by recalling some basic definitions concerning  Feynman
graphs to set up the notations that are going to be used here.
A Feynman graph $G$ is a collection of vertices ${\cal V}_G =
\{V_1,....,V_m\}$ and lines ${\cal L}_G = \{ l_1,....,l_\ell \}$
 associated with a specific term in the
perturbation series of a Green function, and it maps the arguments of
the Green function into points or simple vertices on a plane, and each
propagator into a line (an oriented line in the case of fermion
theories) connecting them. The form of the amplitude corresponding to
the Feynman graph is
$$
G \sim \prod_{V_i \in {\cal V}_G}  \chi (V_i)
\prod_{l_j \in {\cal L}_G} \Delta_{l_j}, \neq \label\amplitude
$$
where $\chi(V_i)$ is the vertex part that carries information about
the interaction, and $\Delta_{l_j}$ is the propagator of the theory
associated to the line $l_j$.

A generalized vertex $U$ of a Feynman graph $G$ is a subgraph of $G$
containing a set of vertices of $G$ together with {\it all} the lines
connecting them.
Given a specific generalized vertex $U$ of a scalar theory its overall
degree of divergence is given through the following formula
$$
\omega(U) = \sum_{conn.} \left (2 \left(\frac{d}{2}-1 \right)
\right) - d(m-1),
\neq \label\degree
$$
where $d$ is the space-time dimension, $m$ the number of simple vertices
that $U$ contains, and the sum is extended over all the lines which
connect two simple vertices.
If $\omega(U) \geq 0$, then $U$ is superficially divergent.

A Feynman graph is connected if it is not the union of two disjoint
diagrams, and is one-particle and irreducible (1PI), if it is
connected and stays so after the removal of any one line. If it is not,
it is called one-particle and reducible (1PR).

We first analyze 1PI graphs which are primitively divergent, that
is, with no subdivergences.
 Consider then a primitively divergent 1PI graph which has two
external vertices
and $\ell$ lines connecting them. The propagator part of the amplitude
is then $$
f (x,y) = \prod_{\ell} \Delta (x-y).
\neq \label\twopbare
$$
One can set $y=0$ due to translational invariance.  DR prescribes
to write $f$ as derivatives, more specifically laplacians
in the case of a scalar theory,
 of a less singular function.
In general
$$
f(x)= \sq^{(N)} A (x),
\neq \label\gtwopdi
$$
where $N$ specifies the number laplacians in the differential operator,
and it is given by the minimun value that satisfies
$$
\omega(A) = \omega (G) -2N <0,
\neq
$$
so that the degree of divergence of $A$ is negative. Then $A$
has a well-defined Fourier transform. For
any sensible renormalizable quantum field theory this number is not
bigger than 2.

The subtraction T-operator for the two-point graph
is defined as
$$
T_{x,0} f(x,0) \equiv f (x,0) - \sq ^{(N)} A(x),
\neq \label\Ttwop
$$
so the renormalized factor is
$$
f^{ren}(x,0) = \left ( 1 - T_{x,0} \right) f(x,0).
\neq
$$

Let us now consider a primitively divergent graph depending on three
vertices such that the divergence
arises when the three points coincide and there is no other partial
subdivergence. We denote by $f(x,y,0)$ this general three point
function. We can use the same strategy used in the three-point example
we studied but let us present a more rigorous way to proceed. First, one
should analyze
$$
\lim_{x \rightarrow 0} \int d^4y \,\varphi(y)\vartheta(y)f(x,y,0),
\neq \label\divthreep
$$
where $\varphi(y)$ is a test function and $\vartheta$ is defined by
$$ \eqalign{
{\rm if} \quad \vert y \vert \le R \quad &{\rm then} \quad
\vartheta(y)=1, \cr
{\rm if} \quad \vert y \vert > R \quad &{\rm then} \quad
\vartheta(y)=0, \cr }
$$
and $R$ is an arbitrary vanishing distance.
Let us rescale the integration variable, $y=\vert x \vert s$
and expand $\varphi(\vert x \vert s)$ in Taylor series around $s=0$.
The degree of divergence
of $f$ will tell us how many orders of the expansion should be kept to
obtain its divergent behavior. If $f$ is logarithmically divergent, the
only term of the Taylor expansion that we need is the first, so
$$
\varphi(0) \int d^4y \,\vartheta(y)f(x,y,0) \equiv \varphi(0) F(x).
\neq \label\defF
$$
 These manipulations are possible since
distributions are  linear and continous mappings on the space of test
functions. That is, if a sequence of test functions $\varphi_k$
converges to $\varphi$ so that the support of $\varphi$ is included in
the union of the supports of the set of $\varphi_k$, then the
sequence given by the action of a distribution $f$ on the $\varphi_k$
converges to the action of $f$ on $\varphi$. The limit $x\to 0$ is only
used to truncate the series in order to isolate only the power
counting divergent terms. This process of regularization is well known
in the mathematical theory of distributions (see for instance ref.
\gouyon ) and can always be applied to tempered singularities, such as
the ones we encounter in the ultraviolet regime of quantum field
theories.

In the limit
$x \rightarrow 0$ we find that $f(x,y,0)$ is a distribution in $y$
with support in $y=0$ that is, a delta function or derivatives of
the delta function, whose coefficient
 $F(x)$ has no Fourier transform. The function $F(x)$ can now be
treated using
the two-point procedure described above. The three-point problem has
been reduced to a two-point one.
We can now define the $T$-subtraction operator for the three-point
graph, which is
 $$
T_{x,y,0} f(x,y,0) \equiv \delta (y) \left(F (x) - \sq
^{(N')} B(x) \right).
\neq  \label\tres
 $$
Therefore, the renormalized factor is obtained carrying out the
operation
$$
f^{ren} (x,y,0) = \left( 1 - T_{x,y,0} \right) f (x,y,0).
\neq
$$

 This procedure can be easily systematized to treat any
n-point primitively divergent graph, so in general
$$
T_{x_1,....,x_{n-1},0} f(x_1,....,x_{n-1},0) \equiv
\delta(x_2 ) ..... \delta(x_{n-1}) \left( F(x_1)- \sq^{(N)} A(x_1)
\right).
\neq \label\npoint
$$
where the subtraction has been reduced to the one of a two-point
function. Note that this property only holds for theories with only
local primitive divergences.
Here $F(x_1)$ is obtained by studying the divergent behavior of
$f(x_1,....,x_{n-1},0)$  when $x_1 \rightarrow 0$, proceeding in the
way that was explained above.
The renormalized factor is
$$
f^{ren}(x_1,....,x_{n-1},0)  =  (1 - T_{x_1,....,x_{n-1},0})
f(x_1,....,x_{n-1},0).
\neq
$$

Let us point out that
we wrote Eq.\tres and Eq.\npoint in a compact form considering that in
both cases the graph was logarithmically divergent. In general, for
more divergent graphs,
$T_{x_1,....,x_{n-1},0} f(x_1,....,x_{n-1},0)$ will also include terms
containing derivatives of delta functions ($Tf$ is said to be a
quasilocal operator). We present in III.4 the
renormalization of a quadratically divergent correlation function of a
composite operator to illustrate this fact.
Notice that we have cast the systematic version of differential
renormalization in
the language of subtraction operators which is the most familiar to the
renormalization community. However, the true spirit of differential
renormalization is closer to the ``replacement operation" defined
by Bogoliubov and Shirkov \bogoliubov. Nevertheless, we have
decided to use
the more standard language of the $T$ subtraction for the sake of
clarity.

Up to now we have studied how to deal with a primitively
divergent 1PI graph. However,
in general, one  must handle the case where apart from the overall
divergence there are also subdivergences, which can be disjoint, nested
or overlapping \ftnote{
Tadpoles are not going to be considered, since DR
chooses to renormalize them to zero in the massless scalar theories we
are presenting.}. It is clear that one has first to subtract off
subdivergences before finding the overall divergence. The exact way to
do it is dictated by the Bogoliubov's recursion formula.
We are going to recall it here.

Remember that
the criteria to find a Feynman graph $G$ free of UV singularities is
given by the Weinberg convergence theorem: If $\omega(U)<0$ for
every generalized vertex $U$ of $G$, including $G$ itself, then $G$ is
absolutely convergent in Euclidean space \weinberg.
Given a Feynman graph $G$ that contains certain divergent subgraphs
($\omega(U) \geq  0$), the renormalized graph $RG$ is given by

\item{} if \hskip 1cm {$\omega(G)<0$} \hskip 3cm {\rm then} \hskip 3cm
{$RG = \overline{R}G$}

\item{} if \hskip 1cm {$\omega(G) \geq 0$} \hskip 3cm {\rm then}
\hskip 3cm {$RG =(1-T_G) \overline{R}G$}

where
$$
\overline{R}G = 1+  \sum_{\cal P} \prod_{U \in \cal P} (-T_U
\overline{R}U) \prod_{conn} \Delta,
\neq  \label\tururu
$$
and the sum extends over all possible partitions $\cal P$ of $G$ into
generalized vertices; $U$ is a generalized vertex belonging to a certain
partition $\cal P$ of $G$; $\prod_{conn.}$ is taken over all lines which
connect the different sets of the partition, and $\Delta$ is the
propagator; $T$ is the DR subtraction operator which acts on $U$ as
follows,

\item{} if $U$ is simple vertex, $T_U U = U$,

\item{} if $U$ is 1PR, $T_U U= 0$,

\item{} if $U$ is 1PI, $T_U$ acts on the propagator part of $U$ as it
has been described.

This concludes the abstract presentation of our general procedure.
The reader may find the above formulae too arid. We lead them to a
down to earth application of the forest construction equipped with our
subtraction prescription in the examples of Sect. III.

\section{Examples}

In this section we present some more involved examples to illustrate the
DR systematic procedure. For simplicity, we bound our study to graphs
occuring in massless euclidean $\phi^4$ theory in four dimensions.

{\bf 1.-The Cateye}

Our first graph ({\fss Fig 3.a}) is a paradigm of overlapping
divergences. We informally call it  ``cateye". Its bare amplitude is
$$
\Gamma (x_1,x_2,x_3,x_4) = {\lambda^4 \over 4 (4 \pi^2)^6} \delta
(x_1-x_2) \delta (x_3-x_4) f(x_1-x_3) + 2 {\rm -perm},
\neq
$$
where
$$
f(x) = \int d^4u d^4v {1 \over u^2 v^2 (x-u)^2 (x-v)^2 (u-v)^4}.
\neq \label\barecat
$$
The forest of this graph is depicted in {\fss Fig 3.b}. Therefore, its
associated forest formula states on $f$ that
$$
f^{ren} (x) = (1-T_{x,0})(1-T_{x,u,v}-T_{0,u,v})(1-T_{u,v}) f(x).
\neq
$$
The second factor tells how to subtract the divergences corresponding to
the two overlapping regions, $x\sim u\sim v$ and $0\sim u\sim v$.

The action of the operator $(1-T_{u,v})$ on $f(x)$ is to substitute in
\barecat the $1/(u-v)^4$ factor by its renormalized value. We have
$$
(1-T_{u,v})f(x)= -\frac14
\int d^4u d^4v {1 \over u^2 v^2 (x-u)^2 (x-v)^2} \sq {\ln (u-v)^2M^2
\over (u-v)^2 }.
\neq \label\bubcat
$$
According to what is prescribed in the previous section, the
subtraction corresponding to each subdivergent region is
$$
T_{x,u,v}(1-T_{u,v})f(x) = $$ $$-\pi^2 \int d^4u d^4v {1 \over u^2 v^2}
\delta (x-u) \left[ {\ln (x-v)^2 M^2 \over (x-v)^4}
+ \frac18 \sq {\ln^2
(x-v)^2 M^2 + 2\ln (x-v)^2 M^2  \over (x-v)^2} \right], \neq
\label\catsubu $$
$$ T_{0,u,v}(1-T_{u,v})f(x) = -\pi^2 \int d^4u d^4v {1 \over (x-u)^2
(x-v)^2}
\delta (u) \left[ {\ln v^2 M^2 \over v^4} + \frac18 \sq {\ln^2
v^2 M^2 + 2\ln v^2 M^2  \over v^2} \right].
\neq \label\catsubdos
$$
We can now integrate by parts the $\sq$ in \bubcat, in order to obtain a
more suitable expression. Notice that this is an {\it exact} operation.
{}From the distribution point of view, it consists in computing a sort of
convolution of three well defined distributions, namely $1/u^2$,
$1/(x-u)^2$ and $\sq \ln (u-v)^2 M^2 / (u-v)^2$, which is well-defined.
We obtain
$$
(1-T_{u,v})f(x) = $$ $$\pi^2 \int d^4v {\ln v^2 M^2 \over x^2 (x-v)^2
v^4} + \pi^2 \int d^4v {\ln (x-v)^2 M^2 \over x^2 v^2 (x-v)^4} $$ $$
 - \frac14 \int d^4u d^4v {1\over v^2 (x-v)^2} \partial_\mu \left(
{1\over u^2} \right) \partial_\mu \left( {1\over (x-u)^2} \right)  {\ln
(u-v)^2 M^2 \over (u-v)^2 }.
\neq \label\ippcat
$$
The first two integrals display logarithmic divergences in the regions
$v\sim 0$ and $v\sim x$, respectively. However, the third term has no
subdivergences whatsoever.
We perform the subtraction of \catsubu and \catsubdos, which clearly
amounts to replacing the divergent factors in the two first terms of
\ippcat. The remaining integrals, which are all well-defined, can be
computed with more or less technical difficulties (see ref \fjl)
yielding
$$
\pi^2 \int d^4v {\ln v^2 M^2 \over x^2 (x-v)^2
v^4} = \pi^2 \int d^4v {\ln (x-v)^2 M^2 \over x^2 v^2 (x-v)^4}
= \frac12 \pi^4 {\ln^2x^2M^2+2\ln x^2M^2 \over x^4}
$$
and
$$
\frac14 \int d^4u d^4v {1\over v^2 (x-v)^2} \partial_\mu \left(
{1\over u^2} \right) \partial_\mu \left( {1\over (x-u)^2} \right)  {\ln
(u-v)^2 M^2 \over (u-v)^2 } = 2\pi^4 {\ln x^2M^2 + 2 \over x^4},
$$
which imply that
$$
(1-T_{x,u,v}-T_{0,u,v})(1-T_{u,v})f(x) = \pi^4{\ln^2x^2M^2 -4 \over
x^4}.
\neq \label\catsub
$$
Actually, the difference between \bubcat and \catsubu and \catsubdos has
no subdivergences. The manipulations that we perform on \bubcat just
help to exhibit that, in effect, the three point divergence is
equivalent to a two point one. From the final expression we obtain,
\catsub, it is straighforward to deal with the remaining overall
divergence. Using the general expression (B.1), we write
$$
f^{ren} (x) = $$ $$(1-T_{x,0})(1-T_{x,u,v}-T_{0,u,v})(1-T_{u,v}) f(x)
= $$ $$ -\frac{\pi^4}{12} \sq {\ln^3x^2M^2 + 3\ln^2x^2M^2 - 6\ln x^2M^2
\over x^2 }. \neq
$$
We have therefore shown that the original reference \fjl~was treating
the overall divergence problem of this graph in the right manner.
The original result was correct because the manipulations performed on
\bubcat happened to split correctly the two overlapping regions of
divergence.

{\bf 2.-The non-planar three-loop graph}

Our next example is the non-planar three-loop graph occuring in the four
point amplitude ({\fss Fig 4}). This graph is primitively divergent so
its forest is trivial. It only exhibits an overall divergence. This can
be checked on its bare expression for the propagator factors,
$$
f(x,y,z,0)={1 \over x^2 y^2 z^2 (x-y)^2 (x-z)^2 (y-z)^2}.
\neq \label\barenonp
$$
The subtraction will be found by studying the behavior of \barenonp in
the limit $x\to 0$, as a distribution in $y$ and $z$. Let $\varphi
(y,z)$ be a
test function. By performing the analysis that was
described in the previous section, we can see that
$$
\lim_{x\to 0} \int d^4y d^4z \varphi (y,z)
{1 \over x^2 y^2 z^2 (x-y)^2 (x-z)^2 (y-z)^2}
\sim \varphi (0,0) \int d^4y d^4z {1 \over x^2 y^2 z^2 (x-y)^2 (x-z)^2
(y-z)^2}. \neq
$$
The integral can be computed using Gegenbauer polynomial techniques.
It yields
$$
\int d^4y d^4z {1 \over x^2 y^2 z^2 (x-y)^2 (x-z)^2 (y-z)^2} = {6\pi^4
\over x^4} \zeta (3),
\neq
$$
where $\zeta$ is the Riemann zeta function. Remark that none of this
integrals has infrared problems. The subtraction necessary to render
finite this graph is therefore,
$$
T_{x,y,z,0} f(x,y,z,0)= 6\pi^4 \zeta (3) \delta(y) \delta(z) \left[
{1\over x^4} + \frac14 \sq {\ln x^2M^2 \over x^2} \right].
\neq \label\subnonp
$$
The difference between \barenonp and \subnonp is a well-defined
distribution in the three variables $x$, $y$, $z$. However, we do not
attain in this case a closed expression for the renormalized amplitude.

{\bf 3.-A six-loop graph}

The following example is the six-loop graph in {\fss Fig 5}.
After dealing trivially with subdivergences, the product of propagators
yields
$$
f(x,y)={\ln x^2M^2 \over x^2} {\ln y^2M^2 \over y^2} \sq {\ln (x-y)^2M^2
\over (x-y)^2}
\neq
$$
The overall divergence of the graph, in the region $x\sim y\sim 0$, has
still to be cured.  The use of the antisymmetric derivative formula
\diffid to reduce the problem to that of a two-point function
is no longer useful here since the $\sq$ would hit now a $\ln y^2M^2
/y^2$ factor which does not produce a delta. We resort to the systematic
DR procedure and study the behavior of $f(x,y)$ in the limit $x\to 0$.
We find
$$
\lim_{x\to 0} f(x,y) \sim -4\pi^2 \delta(y) {\ln^3x^2M^2 \over x^4}
\neq
$$
The subtraction is therefore,
$$
T_{x,y,0} f(x,y) = -4\pi^2 \delta(y) \left[ {\ln^3x^2M^2 \over x^4} +
\frac1{16} \sq {\ln^4x^2M^2 + 4\ln^3x^2M^2 + 12\ln^2x^2M^2 + 24\ln
x^2M^2 \over x^2} \right]
\neq
$$
Now, the renormalized function, $f^{ren} (x,y) = (1-T_{x,y,0})f(x,y)$
is, by definition, a distribution.

We can perform some exact manipulations in $f(x,y)$ in order to obtain a
closed expression for $f^{ren}$. Let us add and subtract from $f(x,y)$,
the following expression,
$$
f_1(x,y)= {\ln^2x^2M^2 \over x^2} {1\over y^2} \sq {\ln (x-y)^2M^2 \over
(x-y)^2 }
\neq
$$
It can be seen that the difference between $f(x,y)$ and $f_1(x,y)$ is
well-defined since both terms need the same subtraction. The
antisymmetric derivative formula \diffid can now be applied to
$f_1(x,y)$.
The promised closed expression for $f^{ren} (x,y)$ is therefore,
$$
f^{ren}(x,y) =
-{\ln x^2M^2 \over x^2} {\ln {x^2\over y^2} \over y^2} \sq {\ln
(x-y)^2M^2 \over (x-y)^2 }+
\partial_\mu \left( {\ln^2x^2M^2 \over x^2} {1\over y^2}
\lrhup{\partial_\mu} {\ln (x-y)^2M^2 \over (x-y)^2 } \right) $$ $$+
\frac14 \pi^2 \delta (y)  \sq {\ln^4x^2M^2 + 4\ln^3x^2M^2 +
12\ln^2x^2M^2 + 24\ln x^2M^2 \over x^2}.
\neq
$$

{\bf 4.-Composite operators}

The last example illustrates the case of a three point overall quadratic
divergence. Such divergence comes out in the computation of the
three-point function,
$$
< : \phi^4 (x):\ : \phi^4 (y):\ :\phi^2 (0): >,
$$
whose Feynman diagram is shown in {\fss Fig.6}. Once the subdivergence
is cured, we face the overall divergence given by
$$
f(x,y,0) = {1\over x^2} {1\over y^2} \sq \sq {\ln (x-y)^2 M^2 \over
(x-y)^2}.
\neq \label\fcomp
$$
Power counting analysis reveals that \fcomp is quadratically
divergent. To find an equivalent divergence, we follow the usual
strategy and compute
$$
\lim_{x\to 0} \int d^4y \varphi (y)
{1\over x^2} {1\over y^2} \sq \sq {\ln (x-y)^2 M^2 \over (x-y)^2},
$$
where $\varphi (y)$ is a test function. Because of the quadratic
divergence,  we have to keep three terms in the Taylor expansion of the
test function. Finally, we find that $f(x,y,0)$ and
$$
\delta (y) {1 \over x^2} \int d^4y {1\over y^2} \sq \sq {\ln (x-y)^2 M^2
\over (x-y)^2}
-\partial_\mu \delta (y) {1 \over x^2} \int d^4y {y^\mu \over y^2} \sq
\sq {\ln (x-y)^2 M^2 \over (x-y)^2}
$$ $$
+\partial_\mu \partial_\nu \delta (y) {1 \over x^2} \int d^4y {y^\mu
y^\nu \over y^2} \sq \sq {\ln (x-y)^2 M^2 \over (x-y)^2}
$$
have the save divergent behavior in the limit $x\to 0$. Again a three
point problem can be reduced to a two point one, thanks to the locality
of the subtraction to be performed. To evaluate its exact form, we need
to compute the $y$-integrals. Such integrals are understood
as being the convolution of two well-defined distributions. Therefore,
they are also well-defined. The calculation yields
$$
\int d^4y {1\over y^2} \sq \sq {\ln (x-y)^2 M^2 \over (x-y)^2} = -4\pi^2
\sq {\ln x^2M^2 \over x^2},
\neq
$$ $$
\int d^4y {y^\mu \over y^2} \sq \sq {\ln (x-y)^2 M^2 \over (x-y)^2} =
8\pi^2 \partial^\mu {\ln x^2M^2 \over x^2},
\neq
$$ $$
\int d^4y {y^\mu y^\nu \over y^2} \sq \sq {\ln (x-y)^2 M^2 \over (x-y)^2}
= 8\pi^2 \left( \delta^{\mu \nu} {\ln x^2M^2 -2 \over x^2} - 4 x^\mu
x^\nu {\ln x^2M^2 -2 \over x^4} \right).
\neq
$$
We just have to find the proper differential identities to construct the
subtraction. For each term, we have,
$$
S_1 (x,y,0) = -4\pi^2 \delta(y) \left[ {1\over x^2} \sq {\ln x^2M^2
\over x^2} -\frac18 \sq \sq {\ln x^2M^2 \over x^2} \right],
\neq
$$ $$
S_2 (x,y,0) = -8\pi^2 \partial_\mu \delta (y) \left[ {1\over x^2}
\partial_\mu {\ln x^2M^2 \over x^2} + \frac1{16} \partial_\mu \sq
{\ln^2x^2M^2 + \ln x^2M^2 \over x^2}\right],
\neq
$$ $$
S_3 (x,y,0) =
8\pi^2 \partial_\mu \partial_\nu \delta (y) \left[ \left( \delta^{\mu
\nu} {\ln
x^2M^2 -2 \over x^4} - 4 x^\mu x^\nu {\ln x^2M^2 -2 \over x^6} \right)
\right.
$$ $$  \left.
- \frac18 \partial_\mu \partial_\nu {1-2\ln x^2M^2 \over x^2}
- \frac1{16} \delta^{\mu \nu} \sq {\ln x^2M^2 \over x^2} \right].
\neq
$$
The renormalized amplitude is thus found by subtracting the three
contributions above,
$$
f^{ren} (x,y,0) = (1-T_{x,y,0})f(x,y,0) = f(x,y,0) - S_1(x,y,0)
-S_2(x,y,0) - S_3(x,y,0).
\neq
$$

\section{Discussion}

Most of the quantum field theory community is used to
renormalizing Feynman amplitudes
using dimensional regularization and minimal subtraction. The common
procedure is to solve
UV problems by redefining the theory in $d$ dimensions, rather than 4.
This analytical continuation gives meaning to the Feynman amplitudes,
and physical results are obtained once poles are removed. The price to
pay for this success is that, at
an intermediate step, all amplitudes have been changed everywhere often
destroying some symmetries which should hold in the final answers. When
the limit to 4 dimensions is taken after subtracting the poles, one
expects to recover the original amplitudes away from the singularities
plus a smooth extension to  all points of space-time. The advantage of
this method is that formal unitarity is mantained since infinities are
absorbed into the bare parameters of the lagrangian.

Let us emphasize that, in principle, no regularization step should be
needed to correct for the few points in space-time where amplitudes are
ill-defined. This is indeed the main philosophy of the standard BPHZ
renormalization procedure as well as ours. Both methods
guarantee unitarity through the correct combination of subtractions when
subdivergences are present, which is achieved through
Bogoliubov's formula.

The main difference between the two methods is the way the subtraction
is performed. In BPHZ, the core of each singularity is isolated by
expanding in Taylor series
the integrand of loop integrals around some external momenta.
 Then, one plainly subtracts this core
singularity from the initial amplitude. A finite result is
obtained upon computation of the subtracted integral.
Differential renormalization, instead, produces a subtraction that
replaces the core of the singularity by its renormalized version.
The subtraction is
done at the level of the amplitude rather than in an integrand and  the
answer naturally carries a renormalization scale, reflecting
the different ways a  function singular at one point
 can be extended into  a distribution.
 The natural scale in
BPHZ comes from the external momenta, whereas in DR scale invariance is
necessarily broken by the integration constant that comes from
writing a singular function as a derivative of less singular
functions. As a consequence of avoiding subtraction in integrands at
zero  external momenta,
 DR seems to bypass infrared problems as compared to BPHZ
where the treatment of massless theories becomes much more involved
\lowens.

{}From a practical point of view, it is remarkable that using differential
renormalization one can compute explicitly complicated
renormalized amplitudes, e.g.
diagram in {\fss Fig.2}. Using BPHZ, one encounters a finite two-loop
integral which is rather involved. Essentially, coordinate space
computations postpone inner integrations to higher loops and produce
more
compact expressions for the amplitudes at low orders in perturbation
theory.

Let us briefly mention some few issues to complete our presentation of
systematic DR. Even though we restricted our study to massless scalar
theories without derivative couplings, the extension to more complicated
cases is
rather straightforward. Fermion theories and theories with derivative
couplings only differ from the case we studied in the change of the
computation of the overall degree of divergence Eq.\degree for a
generalized vertex, and thus in the order of the differential equations
that have to be solved. The extension to massive theories can be easily
done, since the presence of masses does not alter the UV behavior of
amplitudes. If ones chooses to work in a mass independent
renormalization scheme, then one should take the massless limit of the
amplitudes and proceed in the exact way that has been explained to
locate and cure the UV divergences, using the same kind of simple
differential formulas of the massless theory. In a mass dependent
renormalization scheme some  more complicated differential formulas,
involving the presence of Bessel functions, should be used \massive.
In any case, the systematic procedure we have set up holds for all these
theories. In the presence of symmetries, such as gauge symmetries, these
are kept after the renormalization by imposing that amplitudes
fulfill Ward identities.
These identities are seen in DR as relations among
some of the subtraction scales appearing in different amplitudes. The
case of QED has been thoroughly analyzed to two loops in ref. \qed,
with also a study of the chiral anomaly. In this framework, anomalies
result when the Ward identities overconstraint the values of the
renormalization scales \fjl.

We would like to finish by asserting one more time that, as presented
here,
differential renormalization is tailored as a minimal procedure to make
sense out of a field theory. It i) locates and isolates the core of
the singularity of  a bare amplitude, ii) replaces it with a renormalized
version, which only differs from the bare one by a local term, and
 carries an inherent scale, and iii) keeps unitarity by
organizing the subtractions in Bogoliubov's formula which leads to the
fulfillment of RG equations.
Furthermore, the complications due to the use of a regulator are
avoided.
A crucial test still ahead of DR is its application to the weak sector
of the Standard Model. So far, only a Yukawa model with $\gamma_5$ has
been investigated \cris but, there, $\gamma_5$ plays a passive role.

\vfill
\noindent{\flbf Acknowledgments}

This work has been partially supported by CICYT, EEC through the Science
Twinning Grant SC1000337 and NATO under contract \# CRG-910890. C.M. and
X.V.C. acknowledge the Ministerio de Educaci\'on y Ciencia for an FPI
grant.
We would like to thank D.Z. Freedman, P.E. Haagensen, J. Soto and
R. Tarrach for useful discussions.

\vfill \null \eject

\noindent
{\flbf Appendix A}
\bigskip

In this appendix we present an alternative way to prove that eq.
\icecone and \ice have the same divergent behavior in the region $x\sim
y\sim 0$. We simply check that their difference is ultraviolet finite by
going to momentum space. The respective Fourier transforms are
$$
\int d^4y d^4x \, e^{-iyP} e^{-ixQ}  \left (
-\frac14 {1\over x^2} {1\over y^2} \sq {\ln(x-y)^2M^2 \over (x-y)^2}
\right) = -\pi^2 \int d^4p \, {1\over (p-P)^2}{1\over (p-Q)^2} \ln
p^2/\overline{M}^2, $$ $$
\int d^4y d^4x \, e^{-iyP} e^{-ixQ} \left (
\pi^2 \delta(y) {1\over x^2} {\ln x^2M^2 \over x^2} \right) =
-\pi^2 \int d^4p \, {1\over (p-Q)^2} {\ln p^2/\overline{M}^2 \over p^2},
$$
where $\overline{M} = 2M/ \gamma$, and $\gamma = 1.781072...$ is the
Euler constant. The difference is
$$
\pi^2\int d^4p \, {\ln p^2/\overline{M}^2 \over (p-Q)^2} {P^2-2p\cdot P
\over p^2(p-P)^2}
$$
 and it shows to be power counting finite.

\bigskip \bigskip \noindent
{\flbf Appendix B}
\bigskip

We include in this appendix a useful general differential identity.
$$
{\ln^nx^2M^2 \over x^4}= -{1 \over 4(n+1)} \sq \sum_{k=1}^{n+1} {(n+1)!
\over k!} {\ln^k x^2M^2 \over x^2}.
\eqno(B.1)
$$
This equation reproduces the results from \fjl,
$$
{1\over x^4} = -\frac14 \sq {\ln x^2M^2 \over x^2},
$$ $$
{\ln x^2M^2\over x^4} = -\frac18 \sq {\ln^2x^2M^2+2\ln x^2M^2 \over
x^2},
$$ $$
{\ln^2x^2M^2\over x^4} = -\frac1{12} \sq {\ln^3x^2M^2+3\ln^2x^2M^2+6\ln
x^2M^2 \over x^2}.
$$

\immediate\closeout\rfile
\vfill \eject
{\flbf References} \bigskip
\input refs.aux
\vfill \eject

{\fss
\noindent{\flbf Figure Captions}

\item{Fig.1} The one-loop four-point amplitude in $\lambda \phi^4$.

\item{Fig.2} A contribution to the two-loop four-point amplitude in
$\lambda \phi^4$.

\item{Fig.3.a} The cateye.

\item{Fig.3.b} The forest of the cateye.

\item{Fig.4} The non-planar three-loop graph in $\lambda \phi^4$.

\item{Fig.5} A six-loop graph.

\item{Fig.6} A composite operator three-point function, $<:\phi^4(x):\
:\phi^4(y):\ :\phi^2(0):>$.
}
\end